\documentclass[11pt]{article}
\usepackage[centertags]{amsmath}
\usepackage{amsfonts}
\usepackage{amssymb}
\usepackage{amsthm}
\usepackage{newlfont}
\usepackage{color}
\def \n {\noindent}

\usepackage{fancyhdr}
\begin{document}
\begin{center}

{\color{red}(March 2014)}\\

\quad\\

{\Large \bf On the zeros of some families of polynomials satisfying a three-term recurrence associated to Gribov operator}
\end{center}
\begin{center}
*****
\end{center}
\begin{center}
 {\it Abdelkader Intissar}
\end{center}

\n {\it- Equipe d'Analyse Spectrale, UMR-CNRS no:6134, University of Corsica Pascal Paoli.\\ Faculty of Science and Technology, Quartier Grossetti, 20250 Cort\'e, France.}\\
{\it intissar@univ-corse.fr}, {\it T\'el-Fax: 00 33 (0) 4 95 45 00 33}.\\
\begin{center}
 $\&$\quad\quad\\
 \end{center}
{\it- Le Prador, 129, rue du Commandant Rolland, 13008 Marseille,
France}\\

\abstract{ We consider families of tridiagonal- matrices with diagonal
$\beta_{k} = \mu k$ and off-diagonal entries $\alpha_{k} = i\lambda k\sqrt{k+1}$; $1 \leq k \leq n$, $n \in \mathbb{N}$ and $i^{2} = -1$ where $\mu \in \mathbb{C}$ and $\lambda \in \mathbb{C}$.\\\quad In Gribov theory ([7], A reggeon diagram technique, Soviet Phys. JETP 26 (1968), no. 2, 414-423), the parmeters $\mu$ and $\lambda$ are reals and they  are important in the reggeon field theory. In this theory $\mu$ is the intercept of Pomeron  which describes the energy of dependence of total hadronic cross sections
in the currently available range of energies and $\lambda$ is the triple coupling of Pomeron.\\

The main motive of the paper is the localization of eigenvalues $z_{k,n}(\mu, \lambda)$ of the above matrices which are the zeros of the polynomials $P_{n+1}^{^{\mu,\lambda}}(z)$ satisfying a three-term recurrence :\\

$\left\{\begin{array}[c]{l}P_{0}^{^{\mu,\lambda}}(z) = 0\\\quad\\ P_{1}^{^{\mu,\lambda}}(z) = 1\\\quad \\ \alpha_{n-1}P_{n-1}^{^{\mu,\lambda}}(z) + \beta_{n}P_{n}^{^{\mu,\lambda}}(z) + \alpha_{n}P_{n+1}^{^{\mu,\lambda}}(z) = zP_{n}^{^{\mu,\lambda}}(z);\quad n\geq 1\\
\end{array} \right. $\\

\quad\\

\n If $\mu \in \mathbb{R}$ and $\lambda \in \mathbb{R}$ then the above matrices are complex symmetric, in this case we show existence of complex-valued function $\xi(z)$ of bounded variation on $\mathbb{R}$ such that the polynomials $P_{n}^{^{\mu,\lambda}}(z)$ are orthogonal with this weight $\xi(z)$.\\ }\\
$\overline{\quad \quad \quad \quad \quad \quad \quad \quad \quad \quad \quad \quad \quad \quad \quad \quad \quad \quad \quad \quad \quad \quad \quad \quad \quad \quad \quad \quad \quad \quad \quad \quad \quad \quad \quad \quad \quad \quad \quad \quad \quad \quad}$ \\
\n {\bf Keywords:} Jacobi-Gribov matrices; Non self-adjoint Gribov operator; Bargmann space; Polynomials satisfying a three-term recurrence; Eigenvalue problem; Reggeon field theory.\\
{\bf MR(2010) Subject Classification :} 47B36, 47B39, 33C45.\\

$\overline{\quad \quad \quad \quad \quad \quad \quad \quad \quad \quad \quad \quad \quad \quad \quad \quad \quad \quad \quad \quad \quad \quad \quad \quad \quad \quad \quad \quad \quad \quad \quad \quad \quad \quad \quad \quad \quad \quad \quad \quad \quad \quad}$ \\

 \begin{center}
{\Large {\bf 1. Introduction and some preliminaries results}}
 \end{center}

Usually, quantum Hamiltonians are constructed as selfadjoint operators; for certain situations, however, non-selfadjoint Hamiltonians are also of importance. In particular, the reggeon field theory (as invented by V. Gribov {\bf [7]}) for the high energy behavior of soft processes is governed  in zero transverse dimension by the non-selfadjoint operator:\\

$ \mathbb{H} = \mu a^{*}a + i\lambda a^{*}(a + a^{*})a \hfill { }  (1.1)$\\

Where $a$ and $a^{*}$ are the standard Bose annihilation and creation operators that:\\

$[a, a^{*}] = I \hfill { }  (1.2)$\\

and\\

$\mu$ and $\lambda$ are real numbers ($\mu$ is Pomeron intercept and $\lambda$ is the triple coupling of Pomeron see Gribov {\bf[7]} or the excellent survey of the development of Reggeon theory and its application to hadron interactions at high energies given by Borskov et al.at 2006 in {\bf[3]}.) and $i^{2} = -1$.\\

In Bargmann representation {\bf[2]}, the principal spectral properties of $\mathbb{H}$ have been studied in ref.{\bf[8]}.\\

The mathematical difficulties of this problem come of course from the non-self-adjointness of $\mathbb{H}$. Notice
that this non-self-adjointness is a rather wild one; the word "wild" meaning here that the domains of the adjoint
and anti-adjoint parts are not included in one another, nor is the domain of their commutator.\\

We denote the Bargmann space {\bf[2]} by :\\

$\mathbb{B}$ = $\{\phi: I\!\!\!\!C\longrightarrow  I\!\!\!\!C\, entire ; \displaystyle {\int_{I\!\!\!\!C}}\displaystyle{\mid\phi(z)\mid^{2}}e^{-\mid z\mid^{2}}dxdy < \infty \}$ $\hfill { }(1.3)$\\
\vspace{0.5cm}

The scalar product on $\mathbb{B}$ is defined by\\

$<\phi,\psi> = \displaystyle {\int_{I\!\!\!\!C}}\displaystyle{\phi(z)\overline{\psi(z)}e^{-\mid z\mid^{2}}dxdy}\hfill { }(1.4)$\\

and the associated norm is denoted by $\mid\mid . \mid\mid $.\\

$\mathbb{B}$ is closed in $L_{2}(I\!\!\!\!C,d\mu(z))$ where the measure $d\mu(z) = e^{-\mid z\mid^{2}}dxdy$ and it is closed related to $L_{2}(I\!\!R)$ by an unitary transform of $L_{2}(I\!\!R)$ onto $\mathbb{B}$ given in {\bf[2]} by the following integral transform\\

$\phi(z) = \displaystyle{\int_{I\!\!R}}\,e^{-\frac{1}{2}(z^{2} + u^{2}) +\sqrt{2}uz}f(u)du \hfill { }(1.5)$\\

if $f$ $\in L_{2}(I\!\!R)$ the integral converges absolutely.\\

In $\mathbb{B}$ representation $a$ and $a^{*}$ are  the operators of derivation and of multiplication and Gribov operator $ \mathbb{H} := \mathbb{H}_{\mu,\lambda}$ is defined on its maximal domain by\\

$ \left\{\begin{array}[c]{l}\mathbb{H}\phi(z) = \quad i\lambda z\phi^{''}(z)  + (i\lambda z^{2} + \mu z)\phi^{'}(z)\\ \quad \\ with \quad maximal \quad domain: \\\quad \\
D(\mathbb{H}_{max}) = \{\phi \in \mathbb{B} \quad such \quad that \quad  \mathbb{H}\phi \in \mathbb{B}\}\\ \end{array}\right. \hfill { }  (1.6)$\\

\quad\\

Now we give some elementary properties of the operator $\mathbb{H}$ and some remarks.\\

i) $\mathbb{H}$ has the form $\displaystyle{p(z)\frac{d^{2}}{dz^{2}} + q(z)\frac{d}{dz}}$ with $\displaystyle{p(z) = i\lambda z}$ and $\displaystyle{q(z) = i\lambda z^{2} + \mu z}$ of degree 2 then it is of Heun operator type and its eigenvalue problem on $\mathbb{B}$ ($\mathbb{B} = \mathbb{B}_{0}\bigoplus\{constants\}$ where $\mathbb{B}_{0} = \{\phi \in \mathbb{B}$ such that $\phi(0) = 0\}$
 and zero is eigenvalue of  $\mathbb{H}$ without interest) does not satisfying the classical ordinary differential equation of the form:\\

$\displaystyle{\sigma(z)\phi^{''}(z) + \tau(z)\phi^{'}(z) +\alpha\phi(z)= 0}$\\

where $\sigma(z)$ is a polynomial of degree at most two, $\tau(z)$ is a polynomial of degree
exactly one, and $\alpha$ is a constant..\\

ii) An orthonormal basis of $\mathbb{B}$ is given by $\displaystyle{e_{k}(z) = \frac{z^k}{\sqrt{k!}}}; k = 0, 1, ....\hfill { }  (1.7)$\\

iii)  The action of the operators $a$, $a^{*}$ and $\mathbb{H}$ on the basis $\displaystyle{e_{k}(z)= \frac{z^{k}}{\sqrt{k!}}}; k= 0, 1, .....$:\\

$a(e_{k}) = \sqrt{k}e_{k-1}$ with the convention $e_{-1}=0$\\

$a^{*}(e_{k}) = \sqrt{k+1}e_{k+1}$\\

 and\\

$\mathbb{H}(e_{k}) =  i\lambda(k-1)\sqrt{k}e_{k-1} + \mu ke_{k} + i\lambda k\sqrt{k+1}e_{k+1}$\\

iv) Let $\mathcal{P}$ be space of polynomials then it is dense in $\mathbb{B}$.\\

v) Let  $\mathbb{H}_{\mid_{\mathcal{P}}}$ be the restriction of $\mathbb{H}$ to polynomials space, we can define $\mathbb{H}_{min}$ as the closure of operator $\mathbb{H}_{\mid_{\mathcal{P}}}$ in Bargmann space:\\

$ \left\{\begin{array}[c]{l}\mathbb{H}_{min}\phi= i\lambda z\phi^{''}  + (i\lambda z^{2} + \mu z)\phi^{'} \\ \quad \\
with  \quad domain: \\ \quad \\
D(\mathbb{H}_{min}) = \{\phi \in \mathbb{B}; \exists p_{n} \in \mathcal{P}, \psi \in \mathbb{B}, p_{n}\rightarrow \phi \quad and\quad \mathbb{H}p_{n}\rightarrow \psi\}\\ \end{array}\right. \hfill { }  (1.8)$\\

\quad\\

vi)\quad $\mathbb{H}_{min}^{*} = \mathbb{H}_{\mu,-\lambda}$ with domain $D(\mathbb{H}_{min}^{*}) = D(\mathbb{H}_{max}) \hfill { }  (1.9)$\\

where $\mathbb{H}_{min}^{*}$ is the adjoint of $\mathbb{H}_{min}$.\\

Let $l^{2}(\mathbb{N})$ = $\displaystyle{\{ (\phi_{k})_{k=0}^{\infty}\in \mathbb{C}\quad such \quad that\quad \sum_{k=0}^{\infty}\mid\phi_{k}\mid^{2} < +\infty\}}$ with the inner product:\\

$\displaystyle{< (\phi_{k})_{k=0}^{\infty}, (\psi_{k})_{k=0}^{\infty} > = \sum_{k=0}^{\infty}\phi_{k}\bar{\psi}_{k}}$ $\hfill { }  (1.10)$\\

In the representation $l^{2}(\mathbb{N})$ where the coefficients $\phi_{k}$ define an entire function \\

$\phi(z) = \displaystyle{\sum_{k=0}^{\infty}\phi_{k}e_{k}(z)}$ in Bargmann space we have\\

- $ \left\{\begin{array}[c]{l}(a\phi)_{k}=\sqrt{k}\phi_{k-1}, \phi_{-1}= 0\\ \quad \\ with \quad domain: \\ \quad \\ D(a) = \{(\phi)_{k} \in l^{2}(\mathbb{N}); \displaystyle{\sum_{k=0}^{\infty}k\mid \phi_{k}\mid^{2} < \infty} \}\\ \end{array}\right.\hfill { }  (1.11)$\\

\quad\\

- $ \left\{\begin{array}[c]{l}(a^{*}\phi)_{k} =\sqrt{k+1}\phi_{k+1}\\ \quad \\ with \quad domain: \\ \quad \\ D(a^{*})=\{(\phi)_{k} \in l^{2}(\mathbb{N});\displaystyle {\sum_{k=0}^{\infty}k\mid \phi_{k}\mid^{2} < \infty }\} \\ \end{array}\right. \hfill { }  (1.12)$\\

\quad\\

- $ \left\{\begin{array}[c]{l} (\mathbb{H}\phi)_{k} = i\lambda (k-1)\sqrt{k}\phi_{k-1} + \mu k\phi_{k} + i\lambda k\sqrt{k+1}\phi_{k+1}\\ \quad \\ with \quad domain: \\ \quad \\ D(\mathbb{H}) = \{\phi \in l^{2}(\mathbb{N}); H\phi \in l^{2}(I\!\!N)\}\end{array}\right.\hfill { }  (1.13)$\\

\quad\\

We define $l_{0}^{2}(\mathbb{N}) = \{\phi =(\phi_{k})_{k=0}^{\infty} \in l^{2}(\mathbb{N}); \phi_{0} = 0\}$ and the operators $A$ and $A^{*}$ by\\

$A(\phi) = a(\phi)$ with domain $D(A) = D(a)\cap l_{0}^{2}(\mathbb{N})$\\

$A^{*}(\phi) = a^{*}(\phi)$ with domain $D(A^{*}) = D(a^{*})\cap l_{0}^{2}(\mathbb{N})$\\

and now\\

$\mathbb{H} = \mu A^{*}A + i\lambda
A^{*}(A + A^{*})A$ with domain $D(\mathbb{H}) = \{\phi \in l_{0}^{2}(\mathbb{N}); \mathbb{H}\phi \in l_{0}^{2}(\mathbb{N})\}$\\

In the representation $l_{0}^{2}(\mathbb{N})$ where the coefficients $\phi_{k}$ define an entire function\\ $\phi(z) = \displaystyle{\sum_{k=1}^{\infty}\phi_{k}e_{k}(z)}$ in Bargmann space $B_{0}$, In [11], we have study the class of Jacobi-Gribov matrices with unbounded entries:\\

$(H\phi)_{k} = \alpha_{k-1}\phi_{k-1} + \beta_{k}\phi_{n} +\alpha_{k}\phi_{k+1}, k\geq 2  \hfill { }  (1.14)$\\

with the initial condition\\

$(H\phi)_{1} = \beta_{1}\phi_{1} + \alpha_{1}\phi_{2}$, \\

where\\

$\beta_{k} = \mu k$, and $\alpha_{k} = i\lambda k\sqrt{k+1}$, ( $\mu$ and $\lambda$ are real numbers and $i^{2} = -1$).\\

We will write from now the tridiagonal Jacobi-Gribov matrix associated to Gribov operator as\\

    $\mathbb{H} = $ $\left(\begin{array}{c c c c c c c c} \mu & i\lambda \sqrt{2} & 0 & \cdots\\i\lambda \sqrt{2} & 2\mu & i\lambda 2\sqrt{3} &0 & .\\0 &i\lambda 2\sqrt{3} & 3\mu & i\lambda 3\sqrt{4}& 0 &.\\\vdots & 0 & i\lambda 3\sqrt{4} & 4\mu & * & 0 & .\\\vdots & \ddots & 0 & * & * & * & \ddots & .\\\vdots & \ddots & \ddots & 0 & * & * & * & \ddots\\
    \end{array}\right).$ \\

i.e. \\

-$\left\{\begin{array}[c]{l}\mathbb{H} = (h_{j,k})_{j,k=1}^{\infty}\quad with \quad the \quad elements:\\\quad \\h_{kk} = \mu k = \beta_{k}, h_{k,k+1}=h_{k+1,k}= i\lambda k\sqrt{k+1} = \alpha_{k}; k = 1,2,...\\ \quad \\and \\ \quad \\h_{jk} = 0 \quad for \quad \mid j - k \mid > 1 \\ \end{array}\right.\hfill { }  (1.15)$\\

\quad\\

For $ 1 \leq k \leq n$ we remark that\\

i) The eigenvalues $z_{k,n}(\mu, \lambda)$ of the above matrices are the zeros of the Gribov-Intissar polynomials $P_{n+1}^{^{\mu,\lambda}}(z)$ witch  satisfy a three-term recurrence :\\

$\left\{\begin{array}[c]{l}P_{0}^{^{\mu,\lambda}}(z) = 0\\\quad\\ P_{1}^{^{\mu,\lambda}}(z) = 1\\\quad \\ \alpha_{n-1}P_{n-1}^{^{\mu,\lambda}}(z) + \beta_{n}P_{n}^{^{\mu,\lambda}}(z) + \alpha_{n}P_{n+1}^{^{\mu,\lambda}}(z) = zP_{n}^{^{\mu,\lambda}}(z);\quad n\geq 1\\
\end{array} \right. \hfill { }(1.16)$\\

\quad\\

ii) For $\beta_{n} = 2n + \alpha - 1$; $\alpha < - n $ and $\displaystyle{\alpha_{n} = i\sqrt{n(-n - \alpha)}}$, we have the Laguerre polynomials $L_{n}^{^{\alpha}}(z)$.\\

iii) For $\beta_{n} = 0$ and $\displaystyle{\alpha_{n} = i\frac{1}{2}\sqrt{\frac{-n(2\lambda + n - 1)}{(n+\lambda)(n+\lambda - 1)}}}$; $\lambda < - n$, we have the Ultraspherical polynomials $P_{n}^{^{\lambda}}(z)$.\\

iv) For $\displaystyle{\beta_{n} = \frac{\beta^{2} - \alpha^{2}}{(2n +\alpha + \beta + 2)(2n + \alpha + \beta)}}$; $\alpha < - n, \beta < - n, \alpha + \beta < -2(n + 1)$\\

 and \\

$\displaystyle{\alpha_{n} = i\sqrt{\frac{-4(n + 1)(n + \alpha +1)(n + \beta +1)(n + \alpha + \beta +1)}{(2n+\alpha + \beta +2)^{2}(2n+\alpha + \beta + 1)(2n + \alpha + \beta + 3)}}}$, we have the Jacobi polynomials $P_{n}^{^{(\alpha, \beta)}}(z)$.\\

Now, we describe the contents of this paper, section by section. In Section 2, As in [13] or [14], we use a method of functional analysis which transform the problem of the zeros of Gribov-Intissar polynomials $P_{n}^{^{(\mu, \lambda)}}(z)$ to the equivalent problem of the eigenvalues of  Gribov operator and for $\mu \in \mathbb{C}$ and $\lambda \in \mathbb{C}$, we locate the position of the zeros of these polynomials. In Section 3, for $\mu \in \mathbb{R}$ and $\lambda \in \mathbb{R}$, we give some common  properties for Complex Symmetric tridiagonal matrices associated respectively to $L_{n}^{^{\alpha}}(z)$, $P_{n}^{^{\lambda}}(z)$, $P_{n}^{^{(\alpha, \beta)}}(z)$ and $P_{n}^{^{(\mu, \lambda)}}(z)$ and we show the existence of complex-valued function $\xi (z)$ of bounded variation on $\mathbb{R}$ such that the polynomials $P_{n}^{^{(\mu, \lambda)}}(z)$ are orthogonal with this weight $\xi(z)$.\\

For $\mu = 0$, we have applied some results of the excellent article of B. Simon [15](on the classical moment problem as a self-adjoint finite difference operator) to our Gribov operator, see the references [9] and [10] on the complete indeterminacy of this operator acting on Bargmann space.\\
 For a treatment of the spectral analysis of certain Schr$\ddot{o}$dinger operators associated to classical orthogonal polynomials, we refer the interested reader to the excellent article of Ismail-Koelink [12].\\

 \begin{center}
 {\Large{\bf 2. Localization of zeros of polynomials $P_{n}^{^{(\mu, \lambda)}}$}}
 \end{center}

 Let $\mathbb{B}_{0} = \{\phi \in \mathbb{B}; \phi(0) = 0\}$ and
 $\displaystyle{e_{k}(z) = \frac{z^{k}}{\sqrt{n!}}; k = 1, ....}$ an orthonormal basis of $\mathbb{B}_{0}$ .\\

 Let $\mathbb{V}$ be the shift operator defined by $\mathbb{V}e_{k} = e_{k+1}, k = 1, ....$ and its adjoint is the shift operator $\mathbb{V}^{*}$ defined by $\mathbb{V}^{*}e_{k} = e_{k-1}, k = 1, ....$ and $\mathbb{V}^{*}e_{1} = 0$\\

 Let also $\mathbb{A}$ and $\mathbb{B}$ be the diagonal operators defined by $\mathbb{A}e_{k} = \alpha_{k}e_{k}, k = 1, ....$ and $\mathbb{B}e_{k} = \beta_{k}e_{k}, k = 1, ....$ then\\

$ \mathbb{H} = \mathbb{A}\mathbb{V}^{*} + \mathbb{V}\mathbb{A} + \mathbb{B}$ $\hfill { } (2.1)$\\

For $ 1 \leq k \leq n$ the truncated Gribov operator is complex symmetric tridiagonal matrix defined by:\\

$ \mathbb{H}_{n} = \mathbb{A}\mathbb{V}^{*} + \mathbb{V}\mathbb{A} + \mathbb{B}$; $\mathbb{V}e_{n} = 0$ $\hfill { } (2.2)$\\

{\bf Remark 2.1}\\

Let $\tilde{P}_{n}^{^{(\mu, \lambda)}}(z) = det (\mathbb{H}_{n} -zI)$  then we have:\\

$\displaystyle{\alpha_{1}\alpha_{2}......\alpha_{n}P_{n+1}^{^{(\mu, \lambda)}}(z) = (-1)^{n}det (\mathbb{H}_{n} -zI) = (-1)^{n}\tilde{P}_{n}^{^{(\mu, \lambda)}}(z)}$\\

Let $\mu = \mu_{1} + i\mu_{2}$; $\mu_{1}, \mu_{2} \in \mathbb{R}$, $\lambda = \lambda_{1} + i\lambda_{2};\lambda_{1}, \lambda_{2} \in \mathbb{R} $ and $\delta_{k} = k\sqrt{k+1}$\\

Let  $z_{k,n}(\mu, \lambda) = \mathcal{R}ez_{k,n} + i\mathcal{I}mz_{k,n}$ the zeros of the polynomials $P_{n+1}^{^{\mu,\lambda}}(z)$ satisfying  three-term recurrence (1.16)\\

Then we have the following proposition:\\

\quad\\

{\bf Proposition 2.2}\\

i) $\mid \mathcal{R}e z_{k,n} \mid \leq 2\mid \lambda_{2}\mid Sup_{_{k}}\delta_{k} + \mid \mu_{1}\mid Sup_{_{k}} k = 2\mid \lambda_{2}\mid\delta_{n} + n\mid \mu_{1}\mid $\\

ii) $\mid \mathcal{I}m z_{k,n} \mid \leq 2\mid \lambda_{1}\mid Sup_{_{k}}\delta_{k} + \mid \mu_{2}\mid Sup_{_{k}} k = 2\mid \lambda_{1}\mid\delta_{n} + n \mid \mu_{2}\mid $\\

{\bf Proof}\\

Let $\phi_{k} = $ $\left(\begin{array}{c} \phi_{k,1}\\\phi_{k,2} \\.\\.\\\phi_{k,k}\\\\.\\.\\ \phi_{k,n}\\
    \end{array}\right)$ $\in \mathbb{C}^{n}; 1 \leq k \leq n$ and  $\mathbb{H}_{n}\phi_{k} = z_{k,n}\phi_{k};\quad \mid\mid \phi_{k}\mid\mid = 1$\\

Then we have:\\

$z_{k,n} = < \mathbb{H}_{n}\phi_{k} , \phi_{k} > \hfill { } (2.3)$\\

As $\mu \in \mathbb{C}$ and $\lambda \in \mathbb{C}$, we write $\mu = \mu_{1} + i \mu_{2}; (\mu_{1}, \mu_{2}) \in \mathbb{R}^{2}$ and $\lambda = \lambda_{1} + i \lambda_{2};\\ (\lambda_{1}, \lambda_{2}) \in \mathbb{R}^{2}$ for get:\\

$\beta_{n} = \mu_{1}n + i\mu_{2}n$ and $\alpha_{n} = -\lambda_{2}\delta_{n} + i\lambda_{1}\delta_{n} ; \delta_{n} = n\sqrt{n+1}$\\

Let $\mathbb{A}_{1}e_{k} = - \lambda_{2}\delta_{k}e_{k}$, $\mathbb{A}_{2}e_{k} = - \lambda_{1}\delta_{k}e_{k}$, $\mathbb{B}_{1}e_{k} = \mu_{1}ke_{k}$ and $\mathbb{B}_{2}e_{k} = \mu_{2}ke_{k}$\\

Then\\

$\mathbb{H}_{n} = \mathbb{H}_{n}^{1} + i\mathbb{H}_{n}^{2}$ where\\

$\mathbb{H}_{n}^{1} = \mathbb{A}_{1}\mathbb{V}^{*} + \mathbb{V}\mathbb{A}_{1} + \mathbb{B}_{1}$ $\hfill { }  (2.4)$\\

$\mathbb{H}_{n}^{2} = \mathbb{A}_{2}\mathbb{V}^{*} + \mathbb{V}\mathbb{A}_{2} + \mathbb{B}_{2}$ $\hfill { }  (2.5)$\\

and the relation (2.3) takes the form: \\

$z_{k,n} = < \mathbb{H}_{n}^{1}\phi_{k} , \phi_{k} >  + i < \mathbb{H}_{n}^{1}\phi_{k} , \phi_{k} >\hfill { } (2.6)$\\

Since the operators $\mathbb{H}_{n}^{1}$ and $\mathbb{H}_{n}^{2}$ are selfadjoint,
the inner products $< \mathbb{H}_{n}^{1}\phi_{k} , \phi_{k} > $ and $< \mathbb{H}_{n}^{2}\phi_{k} , \phi_{k} > $ are real and as a consequence,\\

$\mathcal{R}ez_{k,n} = < \mathbb{H}_{n}^{1}\phi_{k} , \phi_{k} > $ and $\mathcal{I}mz_{k,n} = < \mathbb{H}_{n}^{2}\phi_{k} , \phi_{k} > $ which immediately result into the following inequalities:\\

i) $\mid \mathcal{R}e z_{k,n} \mid \leq 2\mid \lambda_{2}\mid Sup_{_{k}}\delta_{k} + \mid \mu_{1}\mid Sup_{_{k}} k = 2\mid \lambda_{2}\mid\delta_{n} + n\mid \mu_{1}\mid $\\

ii) $\mid \mathcal{I}m z_{k,n} \mid \leq 2\mid \lambda_{1}\mid Sup_{_{k}}\delta_{k} + \mid \mu_{2}\mid Sup_{_{k}} k = 2\mid \lambda_{1}\mid\delta_{n} + n \mid \mu_{2}\mid $\\

{\bf Remark 2.3}\\

Let $\mu \in \mathbb{R}$ and $\lambda \in \mathbb{R}$ then, in Bargmann representation, we have the following properties on Gribov operator:\\

i) If $n$ goes to infinity, the matrix $\mathbb{H}_{n}^{1}$ generates a self-adjoint operator (the harmonic oscillator).\\

ii) If $n$ goes to infinity, the matrix $\mathbb{H}_{n}^{2}$ generates a symmetric operator (the cubic Heun Operator), but this operator is not self-adjoint because its spectrum is all $\mathbb{C}$. This specificity is attached to degree 2 of $q(z) = i\lambda z^{2} + \mu z$ in (1.6). See the reference [11] for a systematic study of this case.\\

iii) We note that $\alpha_{n}$ is purely imaginary and $\beta_{n}$ real, then $\mathbb{H}_{n}^{*} = \mathbb{V}\mathbb{A}^{*} + \mathbb{A}^{*}\mathbb{V}^{*} + \mathbb{B} $ and thus $\mathbb{H}_{n}$ is obviously not self-adjoint operator. However, it is complex symmetric operator and in the next section, we show that $\mathbb{H}_{n}$ (respectively $\mathbb{H}$) belongs to an interesting class of operators.\\

From the last proposition, we deduce the following corollary:\\

{\bf Corollary 2.4}\\

If $\mu \in \mathbb{R}$ and $\lambda \in \mathbb{R}$, we have:\\

i) $\mid \mathcal{R}e z_{k,n} \mid \leq  \mid \mu\mid Sup_{_{k}} k =  n\mid \mu\mid $\\

ii) $\mid \mathcal{I} mz_{k,n} \mid \leq 2\mid \lambda_{1}\mid Sup_{_{k}}\delta_{k} = 2\mid \lambda\mid\delta_{n} $\\

For the end of this section, we recall from [11] some properties established on the sequence $\tilde{P}_{n}^{^{\mu,\lambda}}(z)$ (under some conditions on the real parameters $\mu$ and $\lambda$).\\

Let $\mu > 0$ then:\\

i) Two successive polynomials $\tilde{P}_{n-1}^{^{\mu,\lambda}}(z)$ and $\tilde{P}_{n}^{^{\mu,\lambda}}(z)$ can not vanish simultaneously.\\

ii) Let $ x \in \mathbb{R}$ such that $\tilde{P}_{n-1}^{^{\mu,\lambda}}(x) = 0$ then $\tilde{P}_{n}^{^{\mu,\lambda}}(x)\tilde{P}_{n-2}^{^{\mu,\lambda}}(x) > 0 \quad\forall n$\\

iii) If $ x < \mu$ then $\tilde{P}_{n}^{^{\mu,\lambda}}(x) > 0 \quad\forall n \geq 2$\\

iv) If $ x > n\mu$ then $(-1)^{n}\tilde{P}_{n}^{^{\mu,\lambda}}(x) > 0 \quad\forall n \geq 2$\\

v) Let $ x \in [\mu, n\mu]$; $n \geq 2$, if $\tilde{P}_{n-1}^{^{\mu,\lambda}}(x)$ and $\tilde{P}_{n-2}^{^{\mu,\lambda}}(x)$ have same sign on $[\mu, n\mu]$ then on this interval $\tilde{P}_{k}^{^{\mu,\lambda}}(x)$ has this same sign \quad$\forall k \geq n-2$\\

vi) Let $\lambda < \frac{\mu}{2\sqrt{2}}$ and $x_{2k+1}$ the smallest zero of $P_{2k+1}^{^{\mu,\lambda}}(x)$
 on $[\mu, n\mu]; n \geq 2$ \\
then the sequence $x_{2k+1}$ is in $[\mu, x_{2}[$ and it is increasing, where $x_{2}$ is the smallest zero of $\tilde{P}_{2}^{^{\mu,\lambda}}(x)$\\

From the above properties we have obtained in [11] the next fundamental result:\\

{\bf Theorem 2.5} [11]\\

For $ \lambda <\frac{\mu}{2\sqrt{2}}$, the Gribov operator $\mathbb{H}$ has a least real eigenvalue.\\

\newpage

\begin{center}
 {\Large{\bf 3. On Complex Symmetric tridiagonal matrix associated to $P_{n}^{^{(\mu, \lambda)}}$ and orthogonality }}
 \end{center}

The Jacobi-Gribov matrix (1.15): \\

$\left\{\begin{array}[c]{l}\mathbb{H} = (h_{j,k})_{j,k=1}^{\infty}\quad with \quad the \quad elements:\\\quad \\h_{kk} = \mu k = \beta_{k}, h_{k,k+1}=h_{k+1,k}= i\lambda k\sqrt{k+1} = \alpha_{k}; k = 1,2,...\\ \quad \\and \\ \quad \\h_{jk} = 0 \quad for \quad \mid j - k \mid > 1 \\ \end{array}\right.$\\

\quad\\

determines two linear operators in $l_{0}^{2}(\mathbb{N})$ by the formal matrix product $\mathbb{H}\phi$. the first operator is defined in the linear manifold of vectors in $l_{0}^{2}(\mathbb{N})$ with finite support related to the set $\mathcal{P}$ of polynomials in Bargmann space, this operator is densely defined and closable. Let $\mathbb{H}_{min}$ be its closure.\\

 The second operator $\mathbb{H}_{max}$ has the domain $D(\mathbb{H}_{max}) =\{ \phi = (\phi_{n})_{n=0}^{\infty} \in l_{0}^{2}(\mathbb{N}); H\phi \in l_{0}^{2}(\mathbb{N})\}$\\

 it was showed in [8] or in [11] (by another method) that $\mathbb{H}_{max} = \mathbb{H}_{min}$\\

 Let $l_{0}^{2}(\mathbb{N})$ with its usuel scalar product :\\

 $\displaystyle{< \phi, \psi > = \sum_{k=1}^{\infty}\phi_{k}\bar{\psi}_{k}}$ \\

 where\\

 $\phi = (\phi_{1}, \phi_{2}, ......, \phi_{k},.....) \in l_{0}^{2}(\mathbb{N})$ and $\psi = (\psi_{1}, \psi_{2}, ......, \psi_{k},.....) \in l_{0}^{2}(\mathbb{N})$\\

 Let $\displaystyle{\mathbb{J} : l_{0}^{2}(\mathbb{N}) \rightarrow l_{0}^{2}(\mathbb{N})}$;  $\displaystyle{\mathbb{J}(\phi) = \bar{\phi} = (\bar{\phi}_{1},\bar{\phi}_{2}, ......, \bar{\phi}_{k},.....)}$ (conjugaison)\\

 The conjugation $\mathbb{J}$ is an antilinear operator in $l_{0}^{2}(\mathbb{N})$ such that\\

 $\displaystyle{\mathbb{J}^{2}\phi = \phi}$, $\phi \in l_{0}^{2}(\mathbb{N})$ $\hfill { } (3.1)$\\

  and\\

  $\displaystyle{< \mathbb{J}\phi, \mathbb{J}\psi > = <\phi, \psi>}$ ;$\phi \in l_{0}^{2}(\mathbb{N})$, $\psi \in l_{0}^{2}(\mathbb{N})$ $\hfill { } (3.2)$\\

  By using the definitions of complex symmetric matrix (complex symmetric operators) given by Garcia-Putinar in [5] or [6], we deduce obvious that $\mathbb{H}_{n}$ is $\mathbb{J}$-symmetric :\\

  $[\mathbb{H}\phi, \psi]_{J} = [\phi, \mathbb{H}\psi]_{J}$ $\hfill { } (3.3)$\\

  where\\

  $[\phi, \psi]_{J} := < \phi, J\psi >$\\

  Observe that (3.3) is equivalent to \\

$\mathbb{J}\mathbb{H}\mathbb{J} =\mathbb{H}^{*}$ $\hfill { } (3.4)$\\

Now, let $\mathbb{T}$ be an unbounded operator on a separable infinite dimensional Banach space $\mathbb{X}$.\\

\noindent We define the following sets:\\

 \noindent $D(\mathbb{T}) = \{\phi \in \mathbb{X} ; \mathbb{T}\phi \in \mathbb{X}\}$ $\hfill { }(3.5)$\\

 \noindent $D(\mathbb{T}^{\infty}) = \bigcap_{n=0}^{\infty}D(\mathbb{T}^{k})$ $\hfill { }(3.6)$\\

{\bf Definition 3.1}\\

 A linear unbounded densely defined operator $(\mathbb{T}, D(\mathbb{T}))$ on a Banach space $\mathbb{X}$ is said to have a simple spectrum  if the following conditions are met:\\

 1) $\mathbb{T}^{k }$ is closed for all positive integers $k$..\\

 2) there exists an element $\phi \in D(\mathbb{T}^{\infty})$ ;\quad $\overline{Sp\{\phi, \mathbb{T}\phi, \mathbb{T}^{2}\phi, .....\}} = \mathbb{X}$\\

{\bf Lemma 3.2}\\

In Bargmann representation $\mathbb{B}_{0}$, the Gribov operator have a simple spectrum.\\

{\bf Proof}\\

Let $\displaystyle{\{e_{k}\}_{k=1}^{\infty}(z) = \frac{z^{k}}{\sqrt{k!}}}$ be an orthonormal basis of $\mathbb{B}_{0}$ and $\mathbb{H}$ be the Jacobi-Gribov matrix defined by (1.15):\\

$\left\{\begin{array}[c]{l}\mathbb{H} e_{1}= \beta_{1}e_{1} +  \alpha_{1}e_{2}\\ \quad\\ \mathbb{H}e_{k} = \alpha_{k-1}e_{k-1} + \beta_{k}e_{k} +  \alpha_{k}e_{k+1}; \quad k = 2,...\\ \quad \\ where \\ \quad\\\beta_{k} = \mu k ; \quad k = 1, 2,...\\\quad \\ and \\\quad \\\alpha_{k} = i\lambda k\sqrt{k +1}; \quad k = 1, 2,...\\ \end{array}\right.\hfill { }  (3.7)$\\

\quad\\

$\mathbb{S} = Sp \{ \mathbb{H}^{k-1}e_{1}; \quad 0\leq k \leq n \}$,  $ n = 1, 2, ....$.\\

We observe that $e_{1} \in \mathbb{S}$ and suppose that\\

$e_{m} \in Sp \{ \mathbb{H}^{k-1}e_{1};\quad 1\leq k \leq m \}$,  $ 1 \leq m \leq n$.\\

As $\alpha_{n} \neq 0$ ; $n = 1, 2, .....$, then by (3.7) we may write :\\

$\displaystyle{e_{n+1} = \frac{1}{\alpha_{n}}[\mathbb{H}e_{n} - \alpha_{n-1}e_{n-1} - \beta_{n}e_{n}]}$\\

then $e_{n+1} \in Sp \{\mathbb{H}^{k-1}e_{1}; \quad 1\leq k \leq n +1 \}$.\\

By induction we conclude that\\

$e_{m} \in Sp \{\quad \mathbb{H}^{k-1}e_{1};\quad  1 \leq k \leq m  \quad \}$ , $ m =1, 2, ......$ $\hfill { } (3.8)$\\

Therefore $\bar{\mathbb{S}} = \mathbb{B}_{0}$, i.e. the operator $\mathbb{H}$ has a simple spectrum.\\

{\bf Lemma 3.3}\\

 Let $\mathbb{G}$ the Gram matrix associated to Gribov operator $\mathbb{H}$ for the system:\\

$\{\hat{e}_{1},\hat{e}_{1}, ........,\hat{e}_{n},\hat{e}_{n}^{*}\}; n \in \mathbb{N}$.\\

where\\

$\hat{e}_{n} = H^{k-1}e_{1}$ and $\hat{e}_{n} = (H^{*})^{^{k-1}}e_{1}$; $k \geq 1$ $\hfill { } (3.5)$\\

Then\\

the vectors $\{\hat{e}_{1},\hat{e}_{1}, ........,\hat{e}_{n},\hat{e}_{n}^{*}\}$, are linearly dependent and their Gram determinant is equal to zero.\\

{\bf Proof}\\

Set $\mathbb{S}_{m} := Sp \{\quad \mathbb{H}^{k-1}e_{1};\quad  1 \leq k \leq m  \quad \}$ , $ m =1, 2, ......$\\

By (3.8), we see that $e_{1}, e_{2}, . . . , e_{m} \in  \mathbb{S}_{m}, m =1, 2, ......$, and therefore $\{e_{k}\}_{k=1}^{m}$  is an orthonormal basis in $\mathbb{S}_{m} , m =1, 2, ......$\\

 Since $\mathbb{J}e_{k} = e_{k}$ , $k =1, 2, ......$, $\mathbb{J}^{2} = \mathbb{J}$ and $\mathbb{J}\mathbb{H}\mathbb{J} = \mathbb{H}^{*}$ we have\\

i) $\mathbb{J}\mathbb{S}_{m} \subset \mathbb{S}_{m}, m =1, 2, ......$.\\

and\\

ii) $(\mathbb{H}^{*})^{m-1}e_{1} = (\mathbb{J}\mathbb{H}\mathbb{J})^{m-1} e_{1} = \mathbb{J}\mathbb{H}^{m-1}\mathbb{J}e_{1} = \mathbb{J}\mathbb{H}^{m-1}e_{1} \in \mathbb{S}_{m}, m =1, 2, ......$.\\

Therefore vectors $e_{1},\mathbb{H}e_{1}, . . . ,\mathbb{H}^{m-1}e_{1}, (\mathbb{H}^{*})^{m-1}e_{1}$, are linearly dependent and their Gram determinant
is equal to zero.\\

{\bf Theorem 3.4}\\

Let $P_{n}^{^{\mu,\lambda}}$ be the sequence  of polynomials satisfying following three-term recurrence associated to gribov operator:\\

$\left\{\begin{array}[c]{l}P_{0}^{^{\mu,\lambda}}(z) = 0\\\quad\\ P_{1}^{^{\mu,\lambda}}(z) = 1\\\quad \\ \alpha_{n-1}P_{n-1}^{^{\mu,\lambda}}(z) + \beta_{n}P_{n}^{^{\mu,\lambda}}(z) + \alpha_{n}P_{n+1}^{^{\mu,\lambda}}(z) = zP_{n}^{^{\mu,\lambda}}(z);\quad n\geq 1\\
\end{array} \right. $\\

\quad\\
where\\

$\displaystyle{\alpha_{n} = i\lambda n\sqrt{n+1}}$; $ n =1, 2, ....$ and $\displaystyle{\beta_{n} = \mu n}$; $ n=1, 2, ......$\\

then\\

there exists a complex-valued functions $\xi(z)$ of bounded variation on $\mathbb{R}$ such that\\

$\displaystyle{\int_{\mathbb{R}}P_{m}^{^{\mu,\lambda}}(z)P_{n}^{^{\mu,\lambda}}(z)d\xi(z) = \delta_{m,n}}$ \quad $m, n=1,2, ..... \hfill { }  (3.9) $\\

{\bf Proof}\\

Let the sequence of polynomials defined recursively by\\

$\displaystyle{\alpha_{n-1}P_{n-1}^{^{\mu,\lambda}}(z) + \beta_{n}P_{n}^{^{\mu,\lambda}}(z) + \alpha_{n}P_{n+1}^{^{\mu,\lambda}}(z) = zP_{n}^{^{\mu,\lambda}}(z)}$;\quad $n\geq 1
\hfill { }  (3.10)$\\

where\\

$P_{1}^{^{\mu,\lambda}}(z) = 1$ and $P_{2}^{^{\mu,\lambda}}(z) = z - \beta_{1}$\\

set\\

$\displaystyle{P_{n}^{^{\mu,\lambda}}(z) = a_{n-1}z^{n-1} + a_{n-2}z^{n-2} + ...+ a_{j}z^{j} +...+ a_{0}; n \geq 2}$ and $ a_{j} \in \mathbb{C}, 0 \leq j \leq n-1$\\

Comparing coefficients by $z^{n}$ in (3.10), we get\\

$a_{n} = \frac{1}{\alpha_{n}}a_{n-1}$,$n = 1, 2, .....$ with $ a_{0} = -\beta_{1} $\\

By induction, we see that \\

$\displaystyle{a_{n} = (\prod_{k=1}^{n}\alpha_{k})^{-1}a_{0}}$\\

Set $\displaystyle{\mathbb{P}_{n}(z) = \prod_{k=1}^{n}\alpha_{k}P_{n}^{^{\mu,\lambda}}(z)}$\\

Multiplying the both sides of (3.10) by $\displaystyle{\prod_{k=1}^{n}\alpha_{k}}$, we obtain:\\

$\displaystyle{\alpha_{n-1}^{2}\mathbb{P}_{n-1} + \beta_{n}\mathbb{P}_{n} + \mathbb{P}_{n+1} = z\mathbb{P}_{n}}$\\

By theorem 6.4 in [4] there exists  a complex-valued functions $\xi(z)$ of bounded variation on $\mathbb{R}$ such that\\

$\displaystyle{\int_{\mathbb{R}}\mathbb{P}_{m}(z)\mathbb{P}_{n}(z)d\xi(z) = (\prod_{k=1}{n}\alpha_{k})^{2}\delta_{m,n}}$ \quad $m, n=1,2, .....$\\

Therefore we get\\

$\displaystyle{\int_{\mathbb{R}}P_{m}^{^{\mu,\lambda}}(z)P_{n}^{^{\mu,\lambda}}(z)d\xi(z) = \delta_{m,n}}$ \quad $m, n=1,2, .....  $\\

\quad\\

\quad\\

{\bf Remark 3.5}\\

In Bargmann space and for $\mu > 0$, it was showed by Aimar and al. in [1], that the  inverse of non-self-adjoint Gribov operator associated to above matrices belongs $\mathcal{C}_{1+\epsilon}$ for all $\epsilon > 0$ ($\mathcal{C}_{p} ; p > 0$  is the class of Carleman operators). By applying the results of this article, we show (in another paper) that more precisely, this inverse is not in $\mathcal{C}_{1}$ where $\mathcal{C}_{1}$  is the class of nuclear operators or trace operators by showing under some conditions on the parameters $\mu$ and $\lambda$ that  $z_{k,n}(\mu, \lambda) \rightarrow k\mu$ as $n \rightarrow +\infty $.\\

\begin{center}
{\bf{\Large References}}
\end{center}

\n{\bf[1] M.T. Aimar, A. Intissar and J.M. Paoli}, Quelques nouvelles propri\'et\'es de r\'egularit\'e de l'op\'erateur de Gribov, Comm. Math. Phys. 172 (1995) 461-466.\\

\n{\bf[2] V. Bargmann, V.}, On a Hilbert space of analytic functions and an associated integral transform I, Comm. Pure Appl. Math., (1962), 14: 187-214.\\

\n {\bf[3] K.G. Boreskov, A. B. Kaidalov, and O.V. Kancheli}, Strong interactions at high energies in the Reggeon approch, Phys. Atomic Nuclei, 2006, 69(10): 1765-1780.\\

\n{\bf[4] Chihara T.S.}, An introduction to orthogonal polynomials, Mathematics and its Applications, Vol. 13, Gordon and Breach Science Publishers, New York – London – Paris, 1978\\

\n{\bf[5] Garcia S.R.}, Putinar M., Complex symmetric operators and applications, Trans. Amer. Math. Soc. 358 (2006), 1285–1315.\\

\n{\bf[6] Garcia S.R.}, Putinar M., Complex symmetric operators and applications. II, Trans. Amer. Math. Soc. 359 (2007), 3913–3931.\\

\n{\bf[7] N. Gribov}, A reggeon diagram technique, Soviet Phys. JETP 26 (1968), no. 2, 414-423\\

\n{\bf[8] A. Intissar}, Etude spectrale d'une famille d'op\'erateurs non-sym\'etriques intervenant dans la th\'eorie des champs de Reggeons, Comm. Math. Phys., 1987, 113(2): 263-297 (in French).\\

\n{\bf[9] A. Intissar}, Analyse de scattering d'un op\'erateur cubique de Heun dans
l'espace de Bargmann, Comm. Math.Phys. 199, (1998), 243-256.\\

\n{\bf[10] A. Intissar}, On the complete indeterminacy and the chaoticity of generalized Heun's operator in Bargmann space, ArXiv:1401.6560v1 [math.SP] 25 Jan (2014) \\

\n{\bf[11] A. Intissar}, Spectral Analysis of Non-self-adjoint Jacobi-Gribov Operator and Asymptotic Analysis of Its Generalized Eigenvectors, Advances in Mathematics (China), Vol.43, No.x, (2014) doi: 10.11845/sxjz.2013117b\\

\n[12] M.E.H Ismail, E. Koelink, Spectral Analysis of Certain Schr$\ddot{o}$dinger Operators, arXiv:1205.0821v2 [math.CA] 15 Sep 2012\\

\n{\bf[13] C. G. Kokologiannaki, D. Rizos}, On the zeros of polynomials satisfying a three-term recurrence relation with complex coefficients, Journal of Inequalities and Special Functions,
Volume 3, Issue 3 (2012), Pages 29-33\\

\n{\bf[14] E.N. Petropoulou}, On the Complex Zeros of Some Families of Orthogonal Polynomials, Abstract and Applied Analysis, Volume 2010, Article ID 263860, doi:10.1155/2010/263860\\

\n{\bf[15] B. Simon}, The Classical Moment Problem as a Self-Adjoint Finite Difference Operator, Advances in Mathematics 137, 82-203 (1998)\\

\end{document}